# Quality control, data cleaning, imputation

[Dawei Liu](#) [1], [Hanne I. Oberman](#) [2], [Johanna Muñoz](#) [3], [Jeroen Hoogland](#) [3], [Thomas P.A. Debray](#) [3,4]

[1] Biogen Digital Health, Biogen, 225 Binney Street, Cambridge, MA 02142, USA

[2] Utrecht University, Padualaan 14, 3584 CH Utrecht, The Netherlands

[3] Julius Center for Health Sciences and Primary Care, University Medical Center Utrecht, Utrecht University, Heidelberglaan 100, Utrecht, the Netherlands

[4] Health Data Research UK and Institute of Health Informatics, University College London, Gibbs Building, 215 Euston Road, London, NW1 2BE, United Kingdom

## Abstract
This chapter addresses important steps during the quality assurance and control of RWD, with particular emphasis on the identification and handling of missing values. A gentle introduction is provided on common statistical and machine learning methods for imputation. We discuss the main strengths and weaknesses of each method, and compare their performance in a literature review. We motivate why the imputation of RWD may require additional efforts to avoid bias, and highlight recent advances that account for informative missingness and repeated observations. Finally, we introduce alternative methods to address incomplete data without the need for imputation.

# Section 1. Introduction

## 1.1 Quality control

Increasingly often, researchers have access to data collected from the routine clinical practice with information on patient health or the delivery of health care from a variety of sources other than traditional clinical trials (1,2). These data are also known as Real World Data (RWD). Some examples of RWD include administrative databases or clinical registries with electronic healthcare records (EHR), which contain information on patient characteristics, admission details, treatment procedures and clinical outcomes (3).

The generation and collection of RWD is often pragmatic, and limited efforts are made to control the data collection scheme or information flow. The quality of RWD thus can vary dramatically across clinical domains and individual databases (4–8). For example, health care records are often incomplete and may contain information that is inaccurate or even inconsistent with other data sources (9,10). It is therefore imperative that studies involving RWD investigate the nature of recorded information to improve their quality, raise awareness on their strengths and weaknesses, and take these into account to facilitate valid inference on the research question at hand.

Although there is no formal framework to assess the quality of RWD, it is common to focus on at least three domains: accuracy, timeliness and completeness (11). Data accuracy relates to the validity of individual data entries (12). It is typically assessed by examining distributional properties of the observed data (e.g., mean, standard deviation, range) and comparing this information with other sources (e.g., previously published population characteristics). Timeliness refers to the degree to which the available data represent reality from the required point in time. Problems can arise when recorded observations (e.g. taken after surgery) do not adequately reflect the patient's health state at the intended measurement time (e.g. before surgery). Finally, completeness represents the existence and amount of missing data.

In this chapter, we first briefly discuss important preprocessing steps in data quality assurance and quality control (QA/QC). Subsequently, we focus on the handling of missing data. As RWD is typically incomplete when missing values are not handled properly, straightforward analysis will very likely lead to misleading conclusions. As such, there is a strong justification to consider and select appropriate analytical methods for handling missing data.

## 1.2 Data preparation

The analysis of RWD often necessitates multiple preprocessing steps to create a meaningful and analyzable dataset from the raw data. In general, we can distinguish between three types of preprocessing steps: data integration, data cleaning, and data transformation.

The first step is to identify and integrate relevant sources of data (e.g. hospital registries, administrative databases) such that all information of interest becomes available for the studied individuals. These data may, for instance, include information on signs and symptoms, diseases, test results, diagnoses, referrals, and mortality. Sometimes, it is also possible to retrieve information from unstructured data sources including texts, audio recordings, and/or images (REF CHAPTER 8 on text mining). When multiple sources of data are available, it is possible to check for duplicate or inconsistent information across data sources, and thus the accuracy of the data can be assessed. Strategies for data integration are discussed in REF CHAPTER 7 on data integration. Once all relevant data sources have been integrated, it is important to select those individuals that are eligible for the intended analysis. The selection requires the identification of the target population, and is often based on disease status or combinations of information (e.g. morbidity code with relevant prescription or results from a diagnostic test). In addition, it is helpful to define relevant time points, including the starting time (also known as index date or baseline) and endpoint (e.g., the outcome of interest) of the study. Although measurements at other time points can be discarded from the dataset, this information can sometimes be used to facilitate risk prediction or missing data imputation (**Section 6.2**). When repeated measurements are available for one or more variables, they can be formatted using two approaches (13). One approach is to code observations made at different time points as separate columns, leading to a so-called "wide format". This approach works well when the repeated measurements occur at regular time intervals, which is rather uncommon for RWD. A second approach is to record repeated information as separate rows, and to include a "time" variable that indicates when the measurements were taken. This approach is also known as the "long format".

As a second step in data preprocessing, it is recommended to inspect the constructed dataset and to generate descriptive summaries such as the mean, standard deviation, range and amount of missing values for each variable (14). This information can be used to assess completeness of the data and to identify outliers with impossible or extreme values. When invalid measurements or recordings are detected, corresponding values can be treated as missing data and subsequently be recovered using imputation methods. Alternatively, in case of extreme but valid values, the analysis may be rendered more robust to outliers by windsorizing (i.e., observations are transformed by limiting extreme values) or trimming (i.e., simply discarding extreme observations). Such methods always cause a loss of information, and their use should be guided by good reasons to reduce the influence of such observations. This will heavily depends on the analysis of interest. For instance, mean and variance measures are heavily affected by outliers, but the median is not affected at all. Unfortunately, it is often difficult to assess the validity of individual measurements. For this reason, researchers may sometimes consider analysis methods that directly account for the (potential) presence of measurement error in the entire dataset during model estimation (REF CHAPTER 9 on measurement error).

Finally, in the last step, data transformations can be performed. For instance, it is sometimes helpful to transform continuous variables (e.g., in line with model assumptions or to improve numerical stability), to re-code categorical variables (e.g., dummy coding to allow unordered and non-equidistant steps between categories), or to collapse multiple variables into an aggregate measure (i.e., data reduction). Further, when the focus of a study is on the development of a prediction model, it is necessary to set up a training and validation set. Although it is common to randomly split the data into two parts, resampling methods have been recommended to make better use of the data in terms of bias and efficiency (REF **Chapter 15 on model evaluation**).

## Section 2. Missing Data

Pre-processing often brings to light that records in some data fields are missing. This requires careful consideration since it may indicate loss of information and almost surely affects the analysis and the subsequent interpretation of findings. The degree to which this is the case primarily relates to the type of missing data. Therefore, first and foremost, it is important to try to understand why data are missing, as this will guide any further processing.

### 2.1 Types of missing data mechanisms

In the broadest sense, there two large groups of missing data mechanisms.

The first group relates to situations where data cannot or should not be measured. For example, it is not possible to assess tumor characteristics or disease severity for healthy patients. Although the absence of any measurements could here be identified and treated as a missing data problem, this strategy should be avoided because it fails to address the fact that no information is actually missing.

The second group arises when variables could have been measured but were not recorded (i.e., information is actually missing). It is, for instance, possible that observations are missing because no measurements were taken or because available measurements were considered invalid or not correctly recorded. Alternatively, it is possible that data collection is complete for individual patients. However, when data are combined across patients or clinical centers, key variables may become incomplete. When trying to understand the consequences of these missing data and to guide the best way forward, it is helpful to distinguish between three mechanisms by which missing data can arise (15): Missing Completely At Random (MCAR), Missing At Random (MAR), and Missing Not At Random (MNAR).

Briefly, MCAR occurs when the probability that a certain type of measurement is missing does not depend on the values of either observed or missing data. This directly implies that missingness is not related to any of the recorded data and that records with missing data do not form any special group. As an example, physical examination records can be lost due to an administrative computer error. There are no measures, either observed or unobserved, that explain missingness for these particular cases: missingness is said to be *completely* at random[1].

In MAR, the probability that a variable is missing differs across records based on the values of observed data. For example, a particular type of diagnostic measure may be ordered more often upon certain blood sample deviations. If these data are indeed missing at random (MAR), this means that the probability that a value is missing is again completely at random *within* subgroups with the same blood sample analysis. That is, after taking observed blood sample measures into account, there is no further information that predicts missingness.

---

[1] The notion of 'completely at random' is intended to mean: not depending on any observed or missing values out of the measures analyzed. Therefore, is does not have to imply that the missing data pattern is totally unsystematic; it may for instance relate to a measure that is not measured and not of interest for the final analysis. Therefore, the definition of MCAR (and equivalently MAR and MNAR) depends on the set of variables of interest.

Lastly, MNAR describes the situation that the probability that a certain type of measurement is missing is associated with *un*observed data. For instance, if certain measures are more often performed in those with a high suspicion of an unfavorable outcome, but this suspicion cannot be derived from other measures that were observed and recorded in the database. An important particular case is where missingness depends on the value of the measure being missing itself. For instance, alcoholics might be less likely to respond to a questionnaire on alcohol intake.

The distinction between these types of missing data mechanisms is helpful when thinking about the inferences one can make based on the observed data only, without modelling the missing data mechanism itself. As it turns out, several methods can obtain unbiased inference when the MAR assumption holds without explicitly modelling the missing data mechanism[2] (See **Section 4**). Although MAR is often a useful and sometimes convenient assumption from a statistical point of view, the analysis of incomplete data will often need to be supplemented by sensitivity analyses that allow for a more complex missingness mechanism (17). Methods for this purpose are discussed in more detail in **Section 6**.

## 2.2 Types of missing data patterns

The manifestation of missing values (regardless of their cause) can be classified into different patterns, each of which requires a different analysis approach. We here focus on common patterns that arise when analyzing RWD.

Real world data are often collected over a period of time and may therefore contain multiple observations for one or more variables. When data are incomplete, it is helpful to distinguish between monotone (e.g., dropout) and non-monotone (intermittent) patterns of missingness (Figure 1). The dropout pattern occurs when a variable is observed up to a certain time-point, and missing thereafter (18). This situation may, for instance, occur when an individual leaves the study prematurely or dies. More generally, a missing data pattern is said to be monotone if the variables can be sorted conveniently according to the percentage of missing data (19). Univariate missing data form a special monotone pattern. The presence of monotone missingness offers important computational savings and can sometimes be addressed using likelihood-based methods (**Section 5.2**). Conversely, the intermittent pattern occurs when an observed value occurs after a missing value. Because the collection of RWD is often driven by local healthcare demands, measurements tend to be unavailable for time points that are of primary interest to researchers. Intermittent patterns of missingness are therefore relatively common for variables that were measured at multiple occasions. In **Section 6.2**, we discuss dedicated imputation methods to address these non-monotone patterns of missingness.

---

[2] In the likelihood and Bayesian paradigm, and when mild regularity conditions are satisfied, the MCAR and MAR mechanisms are ignorable, in the sense that inferences can proceed by analyzing the observed data only, without explicitly addressing the missing data mechanism. In this situation, MNAR mechanisms are nonignorable. Note that in frequentist inference the missingness is generally ignorable only under MCAR (16)

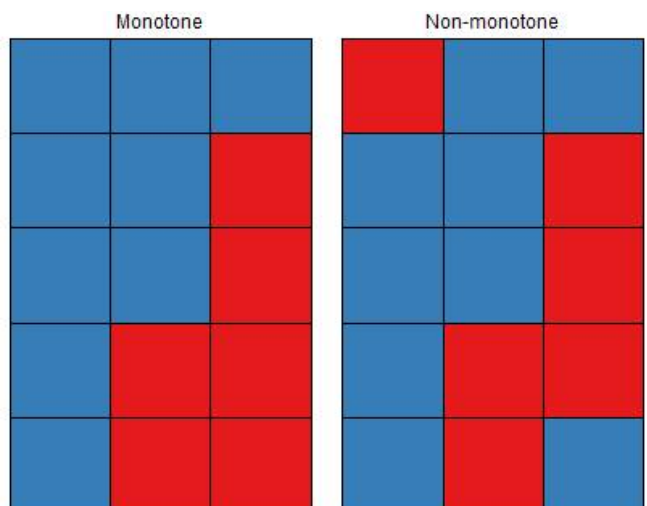

*Figure 1 Illustration of missing data patterns in multivariable data. Each row represents the measurements for a unique patient or timepoint. Columns represent individual variables. Missing values are displayed in red, observed values are displayed in blue.*

Real-world data originating from multiple sources (e.g., hospitals, or even countries) tend to be clustered, with distributions and effects that may differ between clusters. In this context, one can distinguish between data values that are sporadically missing (at least some values available in each cluster) and those that are systematically missing (not measured at all in a particular cluster) (19–21). Systematically missing data are more common when combining routinely collected data from multiple different sources, such as in claims databases. Also, in a pharmacoepidemiologic multi-database studies, there is a high likelihood of missing data because the multiple databases involved may record different variables (22,23). Sporadically missing values often occur and are just the within cluster counterpart of usual missing data. This also leads to the main advantage of dealing with just sporadically missing data. Since at least some information on the joint distribution of the data is available in each cluster, regular missing data methods can be implemented *within* clusters if they have sufficient size. In contrast, more evolved missing data methods that accommodate the clustered nature of the data are necessary to handle systematically missing data. A detailed account of missing data methods designed for clustered data is available elsewhere (19–21).

### 2.3 A bird's eye view on missing data methods

Datasets that are collected in real-world settings are typically large and complex. They are large not only in the sense of the number of individuals, but also in terms of the number of collected variables. At the same time, the structure of RWD also tends to be very complicated. It generally has mixed variable types, containing continuous, categorical and time-to-event variables, some of which could have very sophisticated relationships. It is also common that many variables have missing values and that some variables are incomplete for most individuals. Moreover, when missingness occurs, it is often difficult to determine whether the missing data mechanism is MCAR, MAR or MNAR. Instead, it is very likely that all three missing data mechanisms co-exist in the dataset. The validity of analyses involving RWD will therefore often depend highly on whether missing data were handled appropriately.

Fortunately, several strategies exist to address the presence of missing data. In this chapter, we focus on imputation methods which can address many of the aforementioned challenges. These methods replace the missing values by one (single imputation, see **section 3**) or more (multiple imputation, see **section 4**) plausible values. Imputation avoids the need to discard patient records and separates the missing data problem from the substantive analysis problem (e.g., estimation of a causal effect or predictive model). This implies that imputed data can be analysed using standard methods and software, and as such be directly available for inference (e.g., parameter estimation or hypothesis testing) and the generation of risk predictions. However, as we discuss later in this chapter, single imputation methods are best avoided in most settings because they are not capable of preserving uncertainty about the missing values and their imputation (24). We therefore recommend more advanced approaches that are based on multiple imputation (**section 4**) or avoid imputation altogether (**section 5**). These methods can mainly be applied when data are MCAR or MAR. When the missingness mechanism is MNAR or unknown, additional methods need to be employed (**section 6.1**).

Traditional methods for multiple imputation have been studied extensively in the literature, and are briefly summarized in **sections 4.1 and 4.2**. More recently, numerous imputation methods have also been proposed in the field of machine learning (25). Although these methods tend to be relatively data hungry, they offer increased flexibility and may therefore improve the quality of subsequent analyses (**section 4.3**). To evaluate the potential merit of advanced imputation methods, we embarked on a literature review and focused on imputation methods that are well-suited to handle mixed data types, a large number of both cases and variables, and different types of missing data mechanisms. Briefly, we searched relevant publications on PubMed and ArXiv that describe quantitative evaluations of missing data methods. Initially, we identified 15 relevant papers based on our own experience in the field. These papers compared several statistical and machine learning imputation techniques and were used to inform an active learning literature review. To this purpose, we used the software ASReview, a machine-learning framework that facilitates the screening of titles and abstracts (26,27). To achieve full merit of the framework, a 'stopping criterion' is required–in our case when the software had selected all 15 priorly identified publications. A flow diagram of the review methods is presented in Figure 2. We made use of the following eligibility criteria:

- Inclusion criteria: the paper concerns an evaluation of missing data methods through simulation; the paper matches the search query "(simulation[Title/Abstract]) AND ((missing[Title/Abstract]) OR (incomplete[Title/Abstract]))"; the paper is selected by ASReview before the stopping criterion is reached.

- Exclusion criteria during abstract screening: the paper does not concern an evaluation of missing data methods through simulation; the paper concerns a datatype that deviates from typical EHR data (e.g., imaging data, free text data, traffic sensor data); the paper only concerns (variations of) the *analysis* model, not the *imputation* model; the paper only concerns (variations of) one missing data method.

- Exclusion criteria during full text screening (all of the above, plus): the paper only concerns two missing data methods, one of which is complete case analysis; the paper only concerns single-patient data; the paper only concerns a MCAR missingness mechanism (equivalently, the paper does not concern MAR, MNAR or empirical missingness mechanisms).

After omitting duplicates and removing papers that did not meet the eligibility criteria, we obtained 67 publications. These are listed on [zotero.org/groups/4418459/clinical-applications-of-ai/library](zotero.org/groups/4418459/clinical-applications-of-ai/library).

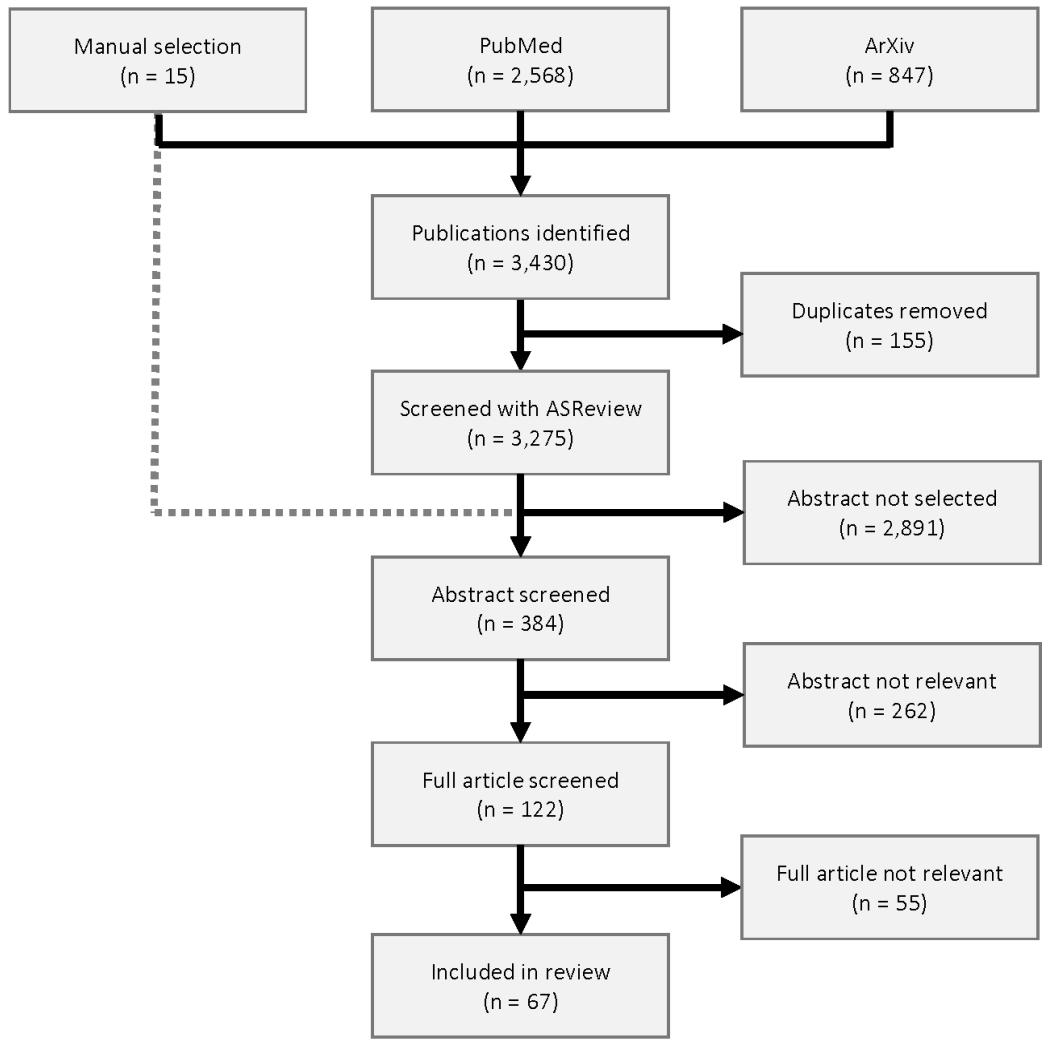

*Figure 2 Flow chart of the literature review to identify quantitative evaluations of missing data methods.*

Based on the aforementioned considerations, we decided to focus on five types of machine learning methods that can be used for imputation: nearest neighbour methods, matrix completion, support vector machines, tree-based ensembles, and neural networks. In the following sections, we briefly introduce each method, discuss its strengths and weaknesses, and provide software implementations. We summarize the main findings from our review in **section 6.3**, offering also a list of recommendations.

## 2.4 Introduction of case study data (MIMIC-III)

The Medical Information Mart for Intensive Care (MIMIC)-III database contains information on 38,597 adults and 7,870 neonates that were admitted to critical care units at Beth Israel Deaconess Medical Center (28,29). Various types of patient-level data are available, including vital signs, laboratory measurements, imaging reports, received treatments, hospital length of stay, and survival. Although many variables were only measured upon admission, temporal data are also available. For instance, there are 753 types of laboratory measurements in MIMIC-III, each with on average 8.13 observations per patient. As illustrated in Figure 3, the missingness rate in MIMIC-III greatly varies between variables and can be as high as 96%.

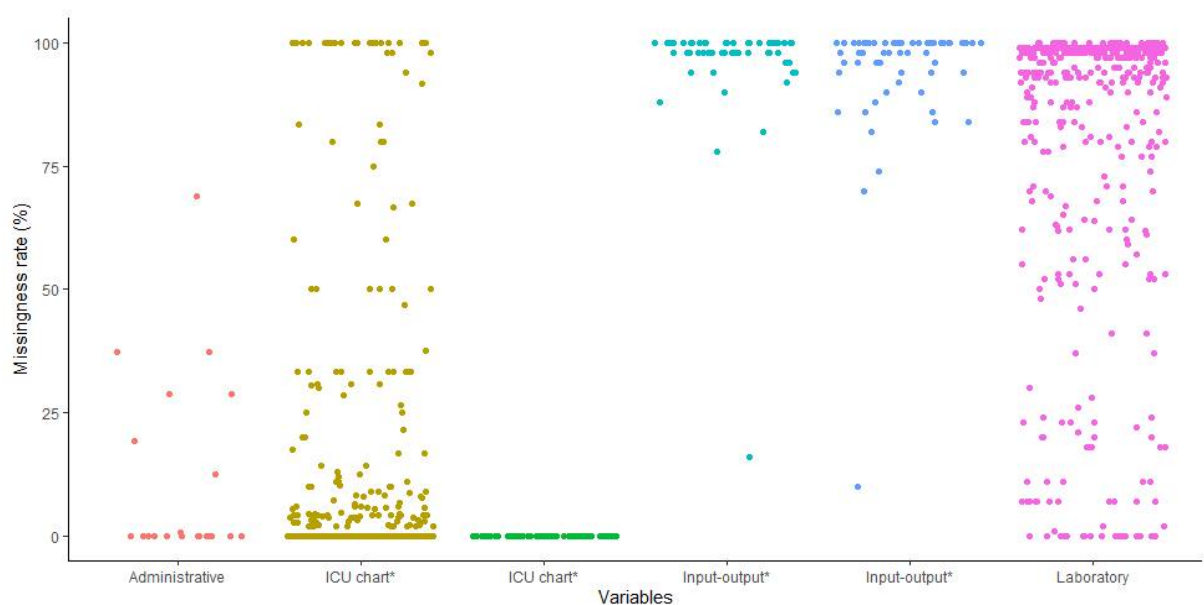

*Figure 3 Visualization of missing data in MIMIC-III (30). Missingness rate is calculated as the proportion of individuals that do not have any observation for a given variable. Administrative variables include demographic data and were not much affected by missing values (e.g., missingness rate for date of birth = 0%). Intensive Care Unit (ICU) chart variables include patient monitoring variables. Input-output variables relate to intake substances (e.g., liquids, medication) and excretions (e.g., urine, fluid from the lungs). Finally, laboratory variables include microbiology results.*
*\* Two different critical care information systems were in place over the data collection period. For this reason, missingness rates for ICU chart and input-output variables are presented as separate categories.*

## Section 3. Single Imputation methods

A common approach to address the presence of missing values is to simply replace them by a plausible value or prediction (31). This approach is adopted by many software packages that implement contemporary machine learning methods. Below, we outline and illustrate three single imputation methods to recover missing systolic blood pressure levels in MIMIC-III.

In single value imputation (SVI), it is widespread to replace missing values of a variable by a convenient summary statistic, such as the mean, median, or mode of the corresponding variable. For example, patients without follow-up data are sometimes assumed to be alive. Similarly, when blood oxygenation levels are incomplete, it is possible to assume that corresponding patients are in perfect health and simply impute a constant that reflects this condition (e.g., 100%). Alternatively, when the health conditions of included patients are suboptimal, it is possible to impute the average of the observed blood oxygenation levels (left graph in Figure 4).

A more advanced approach to generate imputations is to adopt multivariable (e.g., regression or machine learning) models that replace each missing value by a prediction (19). For instance, it is possible to predict blood oxygenation levels in the MIMIC-III database using information on patient age by adopting a regression model (middle graph in Figure 4). As more (auxiliary) variables are used to predict the missing values, the accuracy of imputed values tends to increase (32).

Unfortunately, single imputation methods tend to distort the data distribution because they do not account for sampling variability and model uncertainty (18,19). Because this usually leads to biased inference, single imputation methods are best avoided (31). Their implementation can, however, be acceptable in some circumstances (19). For example, it is possible to add noise to imputed values in order to account for sampling variability (right graph in Figure 4). Also, when applying a prediction model in clinical practice, single imputation methods can greatly facilitate real-time handling of missing values on a case-by-case basis (32,33).

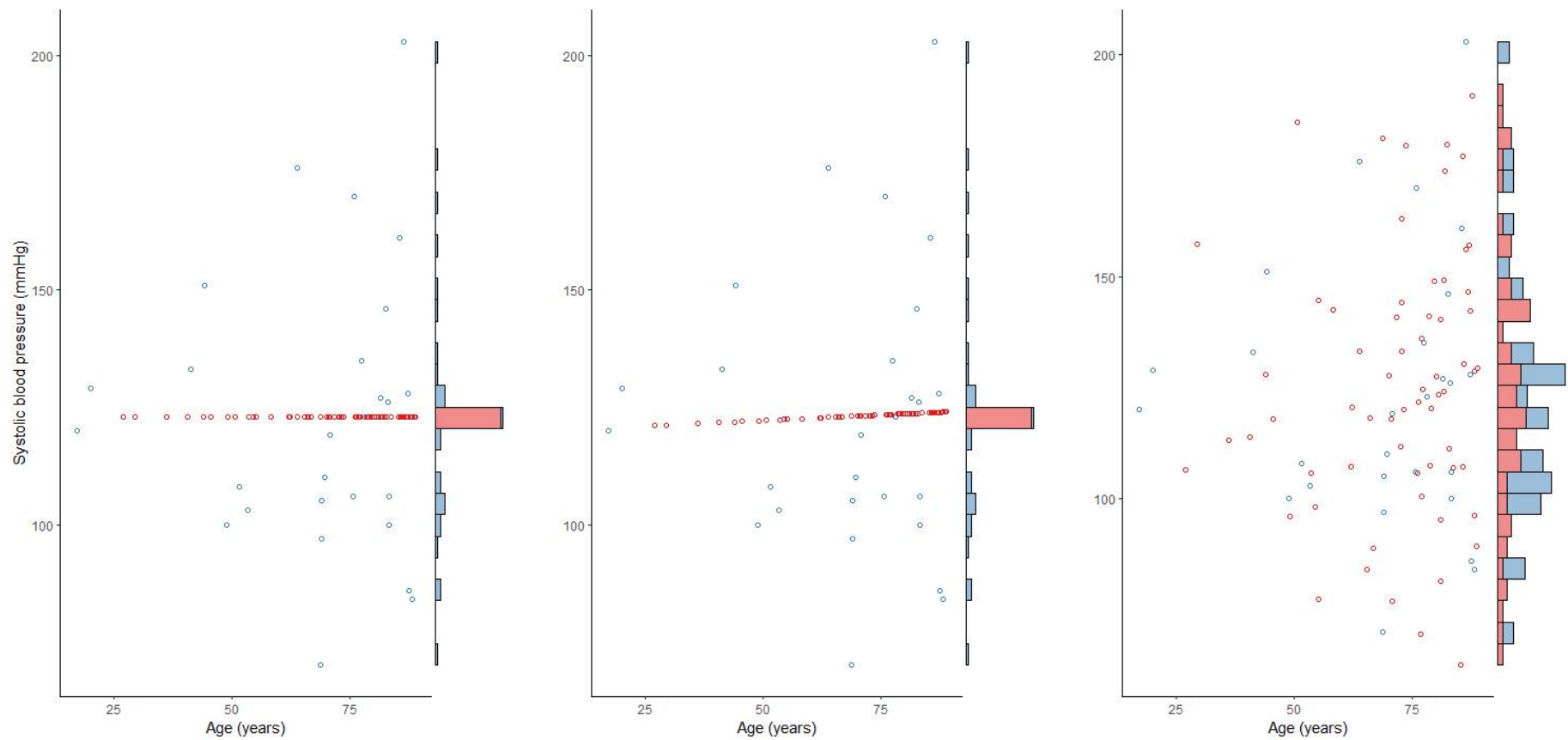

*Figure 4 Illustration of imputation strategies using 100 patients from MIMIC-III. The observed data are displayed in blue and represent the first available measurement for systolic blood pressure after hospital admission. Imputed data are displayed in red, and were generated using mean imputation (left), regression imputation (middle), stochastic regression imputation (right).*

## Section 4. Multiple imputation methods

In general, the preferred approach to address the presence of missing data is to adopt multiple imputation (19,31). In this approach, each missing value in the original dataset is replaced by a set of $m > 1$ simulated values, leading to multiple completed datasets. The entire procedure is illustrated in Figure 5.

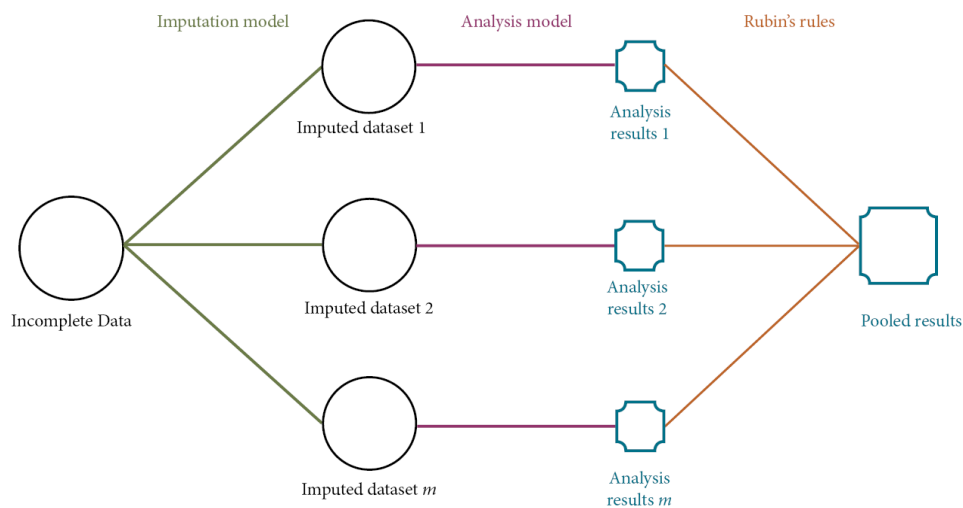

*Figure 5 Scheme of main steps in multiple imputation, adapted from (19).*

The generation of plausible values typically involves modelling the observed data distribution and imposing corresponding parameters on the missing data. A major advantage of multiple imputation is that the extent to which the missing values can accurately be recovered becomes more transparent. The variability of imputed values will be large for variables that cannot adequately be retrieved from the observed data (and vice versa). For example, when temperature measurements are missing for a patient diagnosed with COVID-19 and having symptoms that often coexist with fever, imputed values will have a high probability to indicate the presence of fever. In contrast, fever imputations for a patient with a positive COVID-19 test and only mild disease can be expected to be more variable.

A key challenge in multiple imputation is to generate random samples that are plausible and exhibit an appropriate amount of variability. Conceptually, this can be achieved by generating imputations from a probability distribution. For instance, consider that some patients in MIMIC-III have missing values for age. A simple solution is to approximate the empirical (observed) age distribution, which has a mean value of 65.8 years and a standard deviation of 18.5 years, with a suitable well-known distribution. New values for patient age could then be generated from a normal distribution with the aforementioned characteristics. It may be clear that the aforementioned (univariate) approach does not account for any relation with other variables in the dataset, and thus leads to imputations that are not very plausible. A better approach is to consider the entire (multivariate) distribution of the available data and draw imputations tailored to each patient (34). Here, we discuss two broad strategies to generate personalized imputations: joint modelling imputation and conditional modelling imputation. For the latter, both statistical and machine learning methods can be used. Software implementations are summarized in Figure 6 (Python) and Figure 7 (R).

|  | Imputation method | | | | | | Data type | | Missingness mechanism | |
| --- | --- | --- | --- | --- | --- | --- | --- | --- | --- | --- |
|  | Joint Modelling Imputation | Conditional Modelling imputation | Nearest Neighbor methods | Matrix completion methods | Tree-based ensembles | Neural Networks | Cross-sectional | Longitudinal | M(C)AR | MNAR |
| scikit-learn |  | ✓ | ✓ |  |  |  | ✓ |  | ✓ |  |
| Autoimpute |  | ✓ | ✓ |  |  |  | ✓ |  | ✓ |  |
| statsmodels |  | ✓ |  |  |  |  | ✓ |  | ✓ |  |
| fancyimpute |  |  | ✓ | ✓ |  |  | ✓ |  | ✓ |  |
| matrix-completion |  |  |  | ✓ |  |  | ✓ |  | ✓ |  |
| missingpy |  |  | ✓ |  | ✓ |  | ✓ |  | ✓ |  |
| miceforest |  | ✓ |  |  | ✓ |  | ✓ |  | ✓ |  |
| MisGAN |  |  |  |  |  | ✓ | ✓ |  | ✓ |  |
| MIDASpy |  |  |  |  |  | ✓ | ✓ |  | ✓ |  |

*Figure 6 Python modules for multiple imputation. If an analyst decides to use SVM for imputation, they may need to manually incorporate the algorithm into the imputation procedure. In Python, SVM can be implemented using the scikit-learn library, or using GitHub repositories such as SVMAlgorithm and SupportVectorMachine.*

| | Imputation method | | | | | | Data type | | Missingness mechanism | |
|---|---|---|---|---|---|---|---|---|---|---|
| | Joint Modelling Imputation | Conditional Modelling imputation | Nearest Neighbor methods | Matrix completion methods | Tree-based ensembles | Neural Networks | Cross-sectional | Longitudinal | M(C)AR | MNAR |
| jomo | ✓ | | | | | | ✓ | ✓ | ✓ | |
| pan | ✓ | | | | | | ✓ | ✓ | ✓ | |
| Amelia | ✓ | | | | | | ✓ | | ✓ | |
| jointAI | ✓ | | | | | | ✓ | ✓ | ✓ | |
| mice | | ✓ | ✓ | | ✓ | | ✓ | | ✓ | ✓ |
| miceMNAR | | ✓ | | | | | ✓ | | ✓ | ✓ |
| pan | | ✓ | | | | | ✓ | ✓ | ✓ | |
| HMisc | | ✓ | | | | | ✓ | | ✓ | |
| mitml | ✓ | ✓ | | | | | ✓ | ✓ | ✓ | |
| micemd | | ✓ | | | | | ✓ | ✓ | ✓ | |
| vim | | | ✓ | | | | ✓ | | ✓ | |
| wNNSel | | | ✓ | | | | ✓ | | ✓ | |
| yaImpute | | | ✓ | | | | ✓ | | ✓ | |
| SVDimpute | | | | ✓ | | | ✓ | | ✓ | |
| pcaImpute | | | | ✓ | | | ✓ | | ✓ | |
| softImpute | | | | ✓ | | | ✓ | | ✓ | |
| eimpute | | | | ✓ | | | ✓ | | ✓ | |
| denoiseR | | | | ✓ | | | ✓ | | ✓ | |
| filling | | | | ✓ | | | ✓ | | ✓ | |
| ECLRMC | | | | ✓ | | | ✓ | | ✓ | |
| StructureMC | | | | ✓ | | | ✓ | | ✓ | |
| ROptSpace | | | | ✓ | | | ✓ | | ✓ | |
| missMDA | | | | ✓ | | | ✓ | | ✓ | |
| randomForest | | | | | ✓ | | ✓ | | ✓ | |
| MissForest | | | | | ✓ | | ✓ | | ✓ | |
| randomForestSRC | | | | | ✓ | | ✓ | | ✓ | |
| miForang | | | | | ✓ | | ✓ | | ✓ | |
| missRanger | | | | | ✓ | | ✓ | | ✓ | |
| rMIDAS | | | | | | ✓ | ✓ | | ✓ | |

*Figure 7 Software packages in R for multiple imputation. More detailed information for R packages is available from https://cran.r-project.org/web/views/MissingData.html.*

## 4.1 Joint modelling imputation

A direct approach to consider the entire data distribution is to explicitly specify a parametric joint model for the observed data (35). The parameters of this (imputation) model are estimated from the observed data, and subsequently used to generate imputed values. It is, for instance, common to assume that the observed patient characteristics arise from a multivariate normal model. The mean and covariance can be estimated using Markov Chain Monte Carlo (MCMC) methods and directly be used to draw imputed values that account for individual patient characteristics (33). This approach is also known as multivariate normal imputation (36). Recent work shows that multiple imputation based on more flexible joint models of the data (e.g. allowing for variables of different types, hierarchical structure of the data, or interaction effects) can also be achieved within the Bayesian framework (37,38). Often, it is difficult to identify an appropriate joint model that describes the observed data. Many datasets contain a combination of binary, continuous, categorical, and other data types. These mixed data types usually cannot be described using a multivariate distribution with a well-known density. A common strategy to relax this limitation is to approximate the (multivariate) data distribution by a series of conditional (univariate) distributions which is the focus of the next section.

## 4.2 Conditional modelling imputation

Conditional modelling imputation implies that a separate imputation model is estimated for each incomplete variable (39). For instance, a logistic regression model can be used to describe the conditional distribution of a binary variable (e.g., current smoker). Conversely, a linear regression model can be used to describe the conditional distribution of a continuous variable (e.g., systolic blood pressure). As discussed in **section 4.3**, it is also possible to adopt machine learning models to describe these conditional distributions. Imputed values are then generated by sampling successively from each of the conditional models, which requires an iterative Monte Carlo procedure. This approach is also known as conditional modelling imputation (33), chained equations imputation (40), or fully conditional specification.

## 4.3 Machine learning imputation

Multiple imputation methods often require explicit assumptions about the distribution(s) of the data, including consideration of the potential presence of interactive and non-linear effects. If the imputation model(s) are based on invalid distributional assumptions or fail to incorporate important covariate effects, subsequent analyses can lead to substantial bias (41). For instance, consider that an interaction exists between the age of a patient and their blood test results (which contains missing values). If this interaction is not explicitly accommodated during imputation, its magnitude will be attenuated in the imputed data. Thus, constructing an appropriate imputation model requires consideration of how the imputed data will eventually be used (24). Unfortunately, it is often difficult to predetermine how data will be analyzed, especially when the available data sources were not designed for the intended analysis. It is therefore helpful for imputation models to anticipate certain features of the data (such as interactions, nonlinearities, and complex distributions) without making any specific commitments. Such flexibility can be realized by non-parametric (e.g., nearest neighbor) or semi-parametric models (e.g., neural networks, random forests, or support vector machines) that avoid making distributional assumptions about the observed data. Below, we discuss a selection of common approaches that yield multiple imputed datasets. In general, machine learning methods can be used in two different contexts. One approach is to embed machine learning models in conditional modelling imputation to describe the conditional distribution of a certain variable. For example, missing blood pressure levels could be imputed using a random forest. A second approach is to generate imputed values directly using a dedicated machine learning method, such as matrix completion or adversarial networks.

### Nearest Neighbor methods

Nearest neighbor (NN) methods offer a non-parametric approach to generate imputations without making distributional assumptions. To this purpose, a distance metric is used to determine the relatedness between any two individuals and to identify *neighbors* with complete information for each individual with one or more missing values. Imputation is then achieved by simply copying the observed values from the nearest neighbor (1-NN) or by combining the observed values from *k* nearest neighbors (kNN) into a weighted average (42). Since NN methods generate imputations by (re)sampling from observed data, no special efforts are required to address complex data types. Accordingly, they are often used with incomplete variables that are restricted to a certain range (e.g., due to truncation), skewed, or semi-continuous. To allow for multiple imputed values, NN methods typically determine the distance between two individuals using a random subset of variables, rather than all observed variables (43). Although NN methods can directly be used as a non-parametric imputation approach, they can also serve as an intermediate step in semi-parametric imputation procedures (44). For instance, predictive mean matching combines conditional modelling imputation with NN methods to draw imputations from the observed data (45). It has been demonstrated that NN methods perform well when data are MCAR or MAR (46–48). Although NN methods are simple and easy to implement (49), they strongly depend on the specification of a suitable multivariate distance measure and a reasonably small dimension (since there are fewer near neighbors in high dimensional space). Consequently, the performance of NN methods tends to suffer from high dimensionality problems (50) and declines when *k* is too small or too large. Finally, NN methods do not facilitate the incorporation of MNAR mechanisms, and therefore appear less suitable in RWD.

### Matrix completion methods

Matrix completion methods aim to recover an intact matrix from the dataset with incomplete observations. To this purpose, they decompose the original (high-dimensional) matrix into a product of lower dimensional matrices (51). Missing data are then imputed by identifying an appropriate low-rank approximation to the original data matrix.

For instance, singular value decomposition (SVD) can be used to describe a dataset $X$ with $n$ rows (e.g., patients) and $k$ columns (e.g., variables) by a matrix product $X = UDV'$. In this expression, $D$ is a diagonal matrix with $k$ singular values, $U$ is an $n \times k$ matrix of left singular vectors, and $V$ is an $k \times k$ matrix of right singular vectors. The entries of $D$ are used to scale $U$ and $V$, and therefore describe how much information each singular vector provides to the original data matrix. Recall that the rank of a matrix is the maximal number of linearly independent column vectors or row vectors in the matrix, which is also equal to the number of non-zero singular values of the matrix. By omitting singular values that are close to 0 from $D$ (and omitting the corresponding vectors from $U$ and $V$), the rank of a matrix can be reduced without much loss of information. This, in turn, gives a lower-rank approximation to the original matrix. In case of missing data, the key idea is to find a low-rank approximation that closely fits the observed entries in $X$ from a lower-rank approximation, with the rank sufficiently reduced to fill in the missing parts of $X$.

Other methods that apply matrix completion include (robust) principle component analysis (PCA) and nuclear-norm regularization (51,52). In the latter, the singular values are summarized into a nuclear norm that is optimized using expectation maximization. Matrix completion methods do not make any assumptions about the distribution of the observed data, and can handle high-dimensional data in a straightforward manner. Although their implementation is mainly justified when data are MCAR or MAR, several extensions exist for MNAR situations (53,54). Unfortunately, matrix completion is primarily used for numerical data. For categorical data, mode imputation is generally used. Another limitation is related to the implicit linearity assumption. As rank is a concept for the linear relationship between rows or columns of a matrix, the method does not preserve nonlinear relationships between rows or columns.

*Tree-based ensembles*
Tree-based ensemble methods estimate multiple decision trees on the available data and adopt boosting (e.g., XGBoost) or bagging (e.g., random forests) to combine their predictions. Tree-based ensembles can be applied to mixed data types, do not require distributional assumptions, and naturally allow for variable selection. Moreover, their recursive partitioning operation predisposes to capture nonlinear effects and interactions between variables. Several simulation studies have shown that tree-based ensemble methods can outperform commonly used multiple imputation methods (55–57). We here focus on the use of random forests to generate imputed values, for which at least four different implementations are available (58). In **section 5.3**, we discuss additional approaches for developing random forests without the need for imputation.

The first tree-based approach to handle missing data was proposed by Breiman and is implemented in the R package *randomForest* with the function "rfImpute" (59). It relies on the concept of "proximity" for missing data imputation. Missing values are initially replaced by a simple summary such as their mean or mode, then a forest is constructed and the proximity matrix is calculated. The proximity matrix is a square matrix where each row and column represents a specific individual. Each matrix entry then quantifies the probability that the individuals from the corresponding row and column fall in the same leaf node. The missing value of a particular variable for a specific individual is imputed using an average over the non-missing values of the variable or the most frequent non-missing value where the average or frequency is weighted by the proximities between the case and the non-missing value cases. The process is repeated for each imputed dataset (59).

A second approach termed "on-the-fly-imputation method" was proposed by Ishwara et al. and is implemented in the R package *randomForestSRC* (60). In this method, only observed values are used to calculate the split-statistic when growing a tree. At each node of a tree, when a split decision needs to be made, missing values will be replaced by random observed values within the corresponding subtree. After each node split, imputed values are set back to missing and the process continues until no more splits can be made. Missing data in terminal nodes are then imputed using the mean or mode of out-of-bag non-missing terminal node data from all the trees.

A third approach was proposed by Stekhoven and Buehlmann and is implemented in the R packages *MissForest and missRanger* (55). In this method, missing values are initially imputed using simple methods such as mean or mode. The completed data is then used to construct a forest, which in turn is used to predict the missing values. In contrast to the approach proposed by Breiman, this process of training and predicting iterates until a stopping criterion is met, or until a maximum number of user-specified iterations is reached.

Finally, a fourth approach is to use random forests to approximate the conditional (univariate) distribution of the observed data (61). The chained equations framework is then used to iteratively replace the missing values for each incomplete variable (**section 4.2**). Conditional modeling imputation using random forests has, for instance, been implemented by the function *mice.impute.rf* in the R package *mice* and tends to yield better performance than the three approaches mentioned above (56,62). A major advantage of this approach is that imputed data can be analyzed using any method of choice.

### *Support Vector Machines*
Support Vector Machines (SVM) were developed more than thirty years ago (63,64) and have been successfully used in many real-world applications focusing on classification or prediction. A key building block and also the driving force behind SVM's success is the employment of a kernel function. The kernel function implicitly defines a high-dimensional, or even infinite dimensional feature space (hyperplane), in which data points from different classes could be linearly separated or a continuous response variable could be linearly related to the feature vector. The kernel function needs to be carefully selected, and often takes the form of a Gaussian or polynomial (e.g., when the model should allow for non-linear relations). The most typical scenario for the application of SVM is when all predictors are continuous and when the outcome is binary or continuous. When a predictor variable is categorical, dummy coding needs to be applied. Extensions of SVM are available that can handle categorical or survival outcome data. After the completion of the training process, an SVM generally depends only on a small subset of the original data points, called "support vectors". Although SVM are very powerful in handling high-dimensional data, they are not commonly used for missing data imputation. Possibly, this is because SVM algorithms are very sensitive to noise and less suitable when the sample size is large. For the application of SVM for missing data imputation, no formal statistical software packages were found.

### *Neural Networks*
Neural networks are emerging methods in the field of machine learning and are commonly applied for data generation, feature extraction and dimension reduction. We here discuss two main categories of neural networks that can be used for missing data imputation: autoencoders (AEs) and generative adversarial nets (GANs).

An AE is an artificial neural network specifically designed to learn a representation of the observed data. It typically contains an encoder and a decoder. The encoder maps the original input data to a lower-dimensional representation through successive hidden layers of a neural network. The final layer of an encoder is the output layer, which simply describes the original input layer in a lower dimension (65,66). The decoder then maps the output from the encoder to reconstruct the original input, again through successive hidden layers of a neural network. Unfortunately, standard implementations of AEs require data to be complete, and they may end up learning an identity map (hence perfectly reconstructing the input data when an identity map is used instead of successfully reducing the complexity). To address these problems, several AE variants have been proposed. One approach is to adopt denoising autoencoders (DAE) that corrupt the input data with noise (67). The most common way of adding noise is to randomly set some of the observed input values to zero. This approach can also be applied to incomplete input data, by simply replacing missing values by zero. To facilitate multiple imputation, missing values can be replaced by random samples (68). Further, it is also possible to treat missing values as an additional type of corrupted data, and to draw imputations from an AE trained to minimize the reconstruction error on the originally observed data. This approach has, for instance, been implemented by Multiple Imputation with Denoising Autoencoders (MIDAS) (69). A second extension of AE is to adopt variational autoencoders (VAEs) that learn to encode the input using a latent vector from a probabilistic distribution (70–72). The original data can then be imputed by sampling from the latent posterior distribution.

GANs are another type of neural network that consists of two parts; a generator and a discriminator (73). In an adversarial process, the generator learns to generate samples that resemble the original data distribution, and the discriminator learns to distinguish whether a presented example is original or artificial. The GAN procedure can be extended to allow for the imputation of missing data (74–76). Generative Adversarial Imputation Nets (GAIN) adapt the original GAN architecture as follows (76). The generator learns to model the distribution of the data and to impute missing values accurately. The discriminator then learns to distinguish which values were observed or imputed. The generator's input combines the original input data and a mask matrix that indicates the presence of missing values. Conversely, the input of the discriminator is given by the output of the generator and a hint matrix, which reveals partial information about the missingness of the original data. The discriminator then learns to reconstruct the mask matrix.

### 4.4 Analyzing & combining the imputed datasets

Once multiple imputed datasets have been generated, they can be analyzed separately using the procedure that would have been followed if all data were complete (Figure 5). For example, studies aiming to evaluate a relative treatment effect can perform a regression analysis in the imputed data to estimate an odds ratio adjusted for confounders. From each analysis, one or more parameter estimates (and corresponding estimates of uncertainty) are then obtained and need to be combined. Pooling results across multiple imputed datasets is not trivial and typically requires to consider three sources of uncertainty. In particular, there is estimation error within each imputed dataset (e.g., reflected by the estimated standard errors in each completed dataset), variation due to missing data (reflected by the between-imputation variance of parameter estimates), and uncertainty arising from a finite number of imputations. Although point estimates (e.g., regression coefficients) can simply be averaged across the imputed datasets, the pooling of standard errors requires adopting a series of equations that account for aforementioned sources of uncertainty. These equations are also known as Rubin's rules (34,77,78) and have been implemented in most contemporary software packages.

If pooling is done appropriately, multiple imputation methods yield valid parameter estimates with appropriate confidence intervals. In some situations, however, the implementation of Rubin's rules cannot be justified. For example, an exception arises when data are available for the entire population (79). The application of Rubin's rules also becomes more complicated when imputed datasets are analyzed using non-parametric approaches (e.g., recursive partitioning) or approaches that do not result in the same number of parameters across imputations (e.g., variable selection algorithms) (80–82). In such situations, it may be helpful to avoid imputation altogether.

## Section 5. Non-imputation methods

### 5.1 Complete case analysis

A simple approach to address missing data is to simply remove incomplete records from the dataset. This approach, also known as complete case analysis (CCA), is generally valid but needlessly inefficient under the usually unrealistic MCAR assumption. The adoption of CCA is therefore more appealing when conducting likelihood-based inference under MAR conditions or in datasets where only the outcome is missing (**section 5.2**). Unfortunately, CCA does not offer a solution when estimated models (e.g., for risk prediction or classification) are applied to new patients with incomplete data.

### 5.2 Likelihood-based methods

More advanced approaches to address missing values define a model for the observed data only. For example, survival models can be used to analyze binary outcome variables that are affected by censoring (e.g., due to drop-out). Similarly, multilevel models can be used to analyze repeated outcomes that were measured at arbitrary follow-up times. A special situation arises when missing values only occur for the outcome, as multiple imputation then requires auxiliary variables that are not part of the analysis model to offer an advantage over likelihood-based methods. The adoption of likelihood-based methods is therefore particularly appealing when missingness only depends on covariates that are included in the analysis model (such that missingness is ignorable) (83).

Likelihood-based methods can also be used to address missing covariate values, and often require advanced procedures for parameter estimation (84,85). Although likelihood-based methods tend to be much faster and produce more accurate results than multiple imputation, their applicability is limited to very specific analytical scenarios. Likelihood-based methods may therefore have limited usefulness in RWD, where patterns of missingness can be very complex and additional adjustments may be required to account for other sources of bias (e.g., time-varying confounding).

### 5.3 Pattern submodels

A straightforward alternative to imputation methods is to develop separate models for each missingness pattern. For instance, those individuals for which c-reactive protein (CRP) has been observed contribute to a different model than those individuals for which CRP was not observed. This idea has also been referred to as a pattern submodel approach (86). This type of approach is particularly helpful when the number of missingness patterns is fairly limited with respect to the number of observations, since model development occurs in partitions of the original data. Nonetheless, this is a setting that can be expected to occur quite often RWD. For instance, a whole array of venous blood results, genetics, or imaging data will often be entirely missing or entirely observed. Key benefits of patterns submodels include ease of use (both during development and application) and the fact that it does not rely on assumptions about the missingness pattern. Clear costs include loss of information due to partitioning of the data into missingness patterns (this can be relaxed to allow borrowing of information between patterns, but this invokes the MAR assumption across the patterns for which it is relaxed), and the fact that many models are developed instead of just one. As already noted by Mercaldo and Blume (86), different methods can be envisioned to allow borrowing of information between missingness patterns while retaining some of the robustness with respect to missing data mechanisms, but this is still ongoing research.

## 5.4 Surrogate splits

Surrogate splits is a missing data method that is specific to tree-based methods and was proposed in the context of classification and regression trees (87). The key idea is to not only find the optimal split point when building a tree, but also find second best (or more) split points on variables other than the one providing the optimal split point. This allows using an alternative (surrogate) split variable when the optimal variable is missing. Similar ideas have been proposed throughout tree-based methods research. For instance, instead of finding surrogate splits, the popular XGBoost method (88) finds a default direction for each split point in case the variable to split on is missing. While these methods are easy to apply on any data set with missing values, they have important limitations. For instance, surrogate splits are not able to use information from observed data to infer something about the missing variable. Instead, imputed values are generated conditionally on their position in the tree, which roughly correspond to conditional mean imputation. A more robust approach would be to apply the tree-based methods in multiple imputed data based on flexible methods that preserve more of the data complexities, and subsequently bag the results.

## 5.5 Missing indicator

The indicator method replaces missing values by a fixed value (zero or the mean value for the variable) and the indicators are used as dummy variables in analytical models to indicate that a value was missing. The procedure is applied to each incomplete variable, and can be implemented in any analysis method (e.g., regression, decision trees, neural network). The indicator method allows for systematic differences between the observed and the unobserved data by including the response indicator, and thus to address MNAR. However, its implementation usually leads to biased model parameters and can create peculiar feedback mechanisms between the user of the model (e.g. a clinician) and the model itself (89). For this reason, it is generally discouraged to adopt the missing indicator method for addressing missing data[3].

---

[3] While the details are beyond the scope of this chapter, Mercaldo and Blume (86) describe the implementation of missing indicator methodology in the context of multiple imputation, which does provide unbiased inference and has an interesting relation to the pattern submodels described above.

## Section 6. Imputation of real-world data

Although the principles and methods outlined in **section 4** are primarily designed for imputing missing data in medical studies a clear sampling or data collection design (e.g., an observational cohort study or clinical trial), they can also be applied to incomplete sources of RWD that were not generated under a specific research design. In this section, we discuss two common characteristics of RWD that require more advanced imputation methods and software packages that were discussed in **section 4**. A first challenge is the presence of informative missingness and typically arises when missing data mechanisms are complex and partially unknown. A second challenge is the presence of repeated observations, which occurs when patients are followed for a period of time. Below, we discuss methods that are well suited to address these challenges.

### 6.1 Informative missingness

It is often difficult to determine the exact mechanisms by which missing values occur in RWD. In fact, the distinction between MCAR, MAR and MNAR is a theoretical exercise and all these missingness mechanisms could co-exist in RWD. It is not uncommon that important causes of missingness are not recorded, and missingness in routine healthcare data is often informative (9,22). Unfortunately, traditional imputation methods are not well equipped to address this situation, as they do not distinguish between the observed and missing data distribution.

For example, the CRP test is often ordered when there is suspicion of an infection or an inflammation. Lab results may therefore be missing when elevated levels are deemed unlikely. Although multiple imputation could be used to recover these missing test results from information recorded in the EHR database, this approach is problematic when data on signs and symptoms are unavailable. Similar problems arise when test results are directly linked to their missingness. For instance, it is possible that some patients were referred from another hospital based on their lab results, and therefore did not undergo further testing. In general, when missing data mechanisms depend on unobserved information, the presence of missing values becomes informative about the patient, their physician or even the health care center (90,91).

The plausibility of the MAR assumption (and thus the validity of "traditional" imputation methods) can often be increased by implementing imputation models with auxiliary variables that explain the reasons of missingness during imputation (92). As more patient characteristics are recorded, it becomes less likely that the presence of missing values depends on unobserved information. For instance, when hospital registries only record information on patient age, sex, and blood test results, CRP levels are highly likely to be MNAR when unavailable. Conversely, when information on signs, symptoms, diagnostic suspicions, and other laboratory markers are also recorded, it becomes more likely that these observations explain why CRP is missing. At the very least, it will decrease the influence of MNAR mechanisms.

Unfortunately, the use of auxiliary variables becomes problematic when they are substantially affected by missing values or when they do not strongly predict the presence of missingness. Unfortunately, EHR databases are notoriously prone to prominent levels of missingness, often caused by complex recording processes. For this reason, the imputation of RWD may benefit from more advanced imputation methods that explicitly account for different missing data mechanisms (16). When data are MNAR, it is necessary to model the joint distribution of the data and the missingness through selection, pattern-mixture or shared parameter models (93,94). Selection models factorize the joint distribution into the marginal distribution of the complete data and the distribution of the missingness. As an example, we discuss the Heckman selection model in more detail below (95,96). Conversely, pattern-mixture models separate the marginal distribution for the missingness mechanism and the data distribution conditional on the type of missingness. Essentially, this requires to estimate separate (pattern sub)models for each missingness pattern and to combine their inferences by means of integration. Finally, shared parameter models assume that the data distribution and the missingness indicator are conditionally independent after conditioning on a set of shared parameters or latent variables. This type of model has been successfully applied in settings where the missingness mechanism is related to an underlying process that changes over time. These so-called joint models[4] combine information from a mixed model for a longitudinal outcome and a temporal event model for censoring events with a set of latent variables or random effects.

A common strategy for informative missingness is to directly model the relationship between the risk of a variable being missing and its unseen value (96,97). This strategy is based on the Heckman selection model (95), and can be used to assess and correct potential non-random missingness of outcome data. Briefly, the selection model approach involves two equations to predict the missing value and their availability. Both equations are linked together through their residual error terms, which are modelled using a bivariate (e.g., normal) distribution. The correlation of this distribution is estimated from the available data and indicates to what extent the magnitude of the missing values affects their probability of missingness (i.e., presence of MNAR). A special situation arises when there is no correlation between the error terms, as the Heckman model then generates imputations under the MAR assumption. An important requirement for the implementation of Heckman-type imputation models is the availability of exclusion restriction variables. These variables are related to the probability of missingness, but not to the missing value itself. For example, if younger physicians are more motivated to routinely record data into EHR systems, the age of the treating healthcare professional could be treated as an exclusion restriction variable. Similarly, it is possible that CRP tests are ordered more frequently for patients with a certain healthcare insurance program or socioeconomic background. As discussed, information on missingness mechanisms could also be addressed using traditional imputation methods that adopt auxiliary variables, especially if their inclusion converts MNAR situations into MAR. Indeed, it has been demonstrated that Heckman-selection models perform comparably to traditional imputation methods when missing values do not depend on unobserved information (98). However, Heckman-selection models do not require the MAR assumption and therefore appear more suitable when the missing data mechanisms are unclear. Several simulation studies have demonstrated that Heckman-selection models can greatly decrease bias, even when the proportion of missing data is substantial (96–98).

---

[4] In this context, 'joint' is used to describe models that share a parameter, and is not to be confused with joint models that fully describe a multivariate distribution.

## 6.2. Longitudinal and sequence data

RWD are often collected over a period of time and may therefore contain multiple observations for one or more variables. Traditionally, these data are collected at frequent and regular time intervals. The recorded observations then describe a smooth trajectory that strongly resembles the underlying time process. In RWD, however, there are many challenges as compared to traditional longitudinal data. First, a large number of variables in the dataset are measured over time. For example, the MIMIC-III dataset contains patient medical records from 2001– 2012 and includes thousands of variables with repeated measurements (99). For standard longitudinal or sequence data, the number of variables is generally very small. Second, each variable generally has its own scheme of measurement times, and the measurement interval can be irregular and may even vary across individuals. As illustrated in Figure 6, many clinical variables in the MIMIC-III dataset are affected by irregular measurement times. For standard longitudinal data, all variables typically follow the same scheme of measurement schedule, and for time series data, the measurement interval is fixed and remains the same for the entire series. Third, complex relationships can exist between measurements of different variables at different time points. Finally, missing data can be confounded with the irregularity of measurement schedule, and when missing data do exist, they tend to be informative and the missing rate can be very high for some variables. Due to these challenges, RWD are highly prone to MNAR mechanisms and intermittent patterns of missingness (**section 2.2**). Traditional imputation methods are not capable of handling missing data in longitudinal datasets like EHR. In this section, we therefore discuss advanced imputation methods that are dedicated to longitudinal data. These methods can be used to reconstruct the entire trajectory of longitudinal variables for each distinct individual, but also to recover single observations at particular points in time (e.g., at the startpoint or endpoint of the study).

One approach to address the presence of missing values in longitudinal data is to recover each trajectory separately, using methods designed for time series (TS) reconstruction. Although TS methods were originally designed for the analysis of evenly spaced observations, some methods could also be used when measurement times are irregular (100). It is, for instance, possible to replace the missing values by their respective mean or mode of the repeated measurements. These univariate algorithms are best suited for stationary series (i.e., when statistical properties of the data generation process do not change over time) and should generally be avoided because they tend to introduce bias for non-stationary series. More advanced univariate algorithms for TS imputation may account for trend (i.e., the long-term direction of the data), seasonality (i.e., systematic patterns that repeat periodically), or even certain irregularities (i.e., distribution of the residuals) of the repeated observations (101). These algorithms often rely on moving averages or interpolation methods, and can be satisfactory when the stretches of missing data are short and if the TS is not much affected by noise (102). Last observation carried forward (LOCF) is a special type of interpolation, where the last observed value replaces the next missing observations. Another common example is the use of autoregressive integrated moving average (ARIMA) models, which eliminate autoregressive parts from the TS and can also adjust for seasonality. However, because their implementation can distort the data distribution and their relation with other variables, univariate TS algorithms should be used with caution. Instead, multivariate TS algorithms could be used to create time lagged and lead data, and to include smooth basis functions over time in the imputation model (103).

Simulation studies found that this strategy tends to outperform simple univariate TS algorithms (102). There are several R packages available for missing data imputation in time series. Due to space limitations, we will not list the packages individually, and refer the reader to https://CRAN.R-project.org/view=TimeSeries.

A different class of methods allows borrowing of information across individuals. When repeated measurements are structured in the wide format, time-related variables can be imputed using the methods discussed in **section 4** without the need for further adjustment. For instance, the Sequential Organ Failure Assessment (SOFA) score is widely employed in the daily monitoring of acute morbidity in critical care units (104). It is calculated using information on the patient's respiratory, cardiovascular, hepatic, coagulation, renal and neurological systems, and prone to missing values when some test results are unavailable. For example, most predictors of the SOFA score were affected by missing values in MIMIC-III, with missingness rates ranging from 58.88% to 99.98% (Figure 6). When repeated measurements of the SOFA score are formatted into separate columns with daily observations, corresponding variables can be imputed with joint modelling methods such as JM-MVN (31) or with conditional modelling methods such as FCS-fold (105). Alternatively, recurrent neural networks can be used to capture long-term temporal dependencies without the need for distributional assumptions (30,101,106–108). Also other machine learning methods have been customized to allow for imputation of longitudinal data, including matrix completion and nearest neighbor methods (101). Unfortunately, these methods are not well suited to recover irregularly spaced observations.

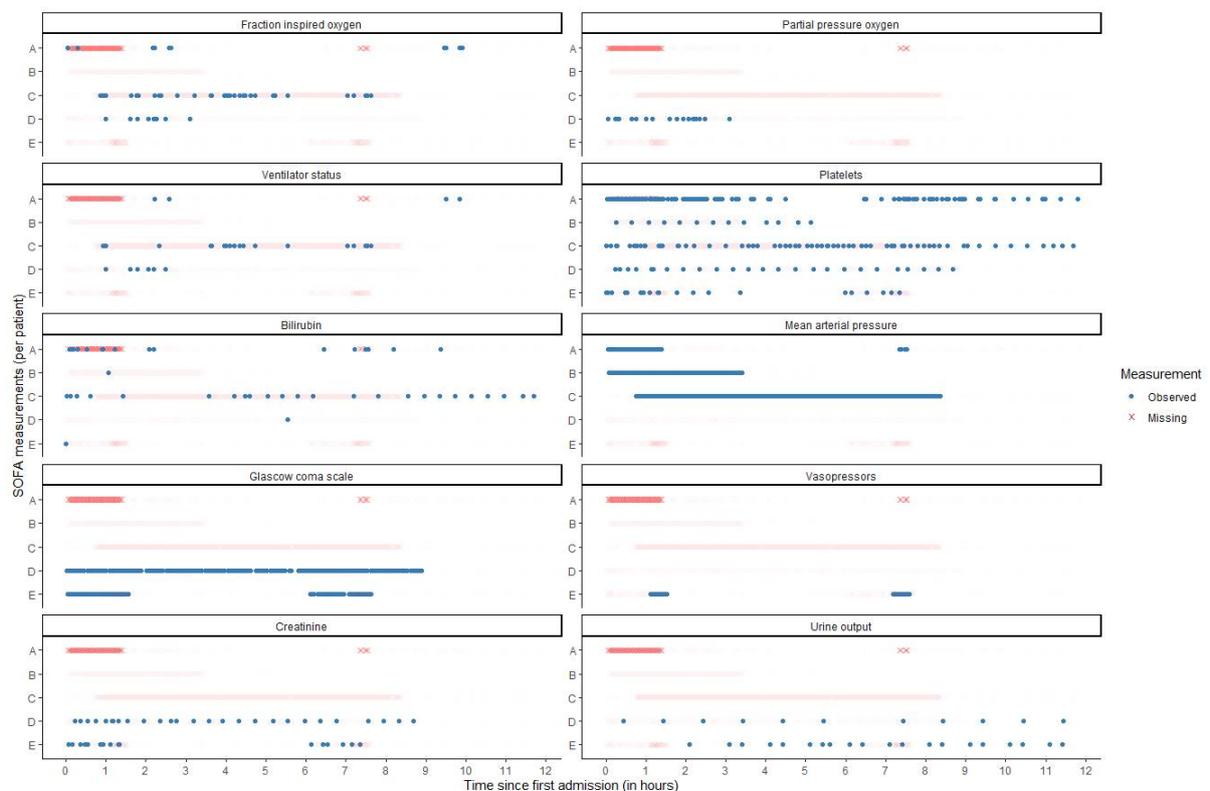

*Figure 8 Illustration of longitudinal data for five patients from MIMIC-III. Repeated measurements are presented for all predictors of the SOFA score. Each point represents a contact moment between the patient and healthcare provider.*

Because RWD are rarely collected at regular time intervals, it is often more helpful to structure sequential observations in the long format. Imputation methods then need to adjust for the time of measurement and the non-independence of observations. This requires to adopt hierarchical (also known as multilevel) models that group related observations, which can be achieved using joint modelling or conditional modelling imputation. A detailed overview of imputation methods for longitudinal data is provided by Huque et al. (109). When adopting multilevel imputation methods, the longitudinal relation of repeated observations can be preserved by including measurement time as an explanatory (possibly random) variable. It is then common to assume a linear relationship for the effect of the time variable. Unfortunately, this approach may become problematic when there is no linear association between the incomplete variable and its predictors, when there is no compatibility between the joint distribution and the full conditional model, or when there is a lack of congeniality between the imputation model and the analysis model. Therefore, it has been proposed to adopt Bayesian substantive-model-compatible methods in which the joint distribution of the variables in the imputation model is specified by a substantive analysis model and an incomplete explanatory variable model (37,110,111). Alternatively, van Buuren proposed the time raster imputation method (13) to convert irregular observations into a set of regular measurements using a piecewise linear mixed model. Initially, the user must specify an ordered set of *k* break times. Next, a B-spline model is used to represent each subject's time points with knots that are given by *k*. This approach then yields a k-column matrix *X*. Finally, the incomplete time-dependent variables are imputed using chained equations with a clustering method, using the reference variables, other time-dependent variables and *X* as predictors in the imputation method for each incomplete variable. More recently, Debray et al. developed conditional modelling imputation methods that adjust for clustering and autocorrelation. These methods were implemented using chained equations and can be used to recover missing observations at arbitrary time points (112). Simulations showed that this approach substantially outperforms simpler imputation methods such as LOCF or rounding, and can also yield valid inferences when longitudinal data are MNAR.

It is also possible to impute longitudinal data using machine learning methods such as recurrent neural networks (RNNs). Although RNNs have been described since 1986 (113), they have rarely been used for longitudinal data analysis until the past decade. Standard RNNs bear many similarities to traditional feedforward neural networks, and can use the output from previous time steps as input for the next time step. In this manner, RNNs offer the ability to handle sequential or time series data. Traditional implementations of RNN cannot process information across many time steps and therefore have a short-term memory. This limitation can be addressed by adopting gated architectures that control the flow of information in the RNN (114). The long short-term memory (LSTM) (115) and the Gated Recurrent Unit (GRU) are common examples of this architecture (116).

Traditional RNNs require that all variables have the same measurement schedule. For this reason, they are not well suited for the imputation and analysis of RWD. In the past few years, there have been tremendous research developments to facilitate the analysis of multivariate sequence data collected with irregular measurement schedules (101). RNN methos can, for instance, be enhanced by adopting adversarial training, attention mechanisms, or multidirectional structures. We here distinguish between three common types of RNN for imputation of longitudinal data. A first type of RNN methods generate multiple imputed datasets, and include Bidirectional Recurrent Imputation for Time Series (117), multi-directional recurrent neural networks (118), and residual neural networks (119). A second, similar type of RNN methods adopt generative adversarial networks to learn the overall distribution of a multivariate time series data and to generate imputed datasets (120). Finally, a third type of RNN methods do not yield imputed datasets, but offer an integrated solution to the analysis of incomplete longitudinal data. To this purpose, they adopt missing indicators ("masks") and/or the time interval between the observed values as input values of the network (30,121–123). To increase the ability to capture long-term relations in the data, these non-imputation methods often adopt a GRU or LSTM architecture. Estimation of aforementioned RNNs is not straightforward and often requires dedicated software packages, which may not always be readily available or easy to use.

## 6.3 Choosing an appropriate imputation method

The selection of an appropriate imputation method will often depend on the ultimate goal of the data analysis. If the goal is to make statistical inferences, such as estimating regression parameters or testing certain hypotheses, it is important that the imputation method provides not only unbiased estimates of parameters of interest, but also unbiased estimates of their associated (co)variance. On the other hand, if the goal of data analysis is to make predictions or classification, a suitable imputation method should be able to maintain the desired prediction or classification accuracy.

As discussed in this chapter, multiple imputation offers a generic solution to handle the presence of missing data. Multiple imputation can be used in both "inference-focused" and "prediction-focused" studies, and can also be used on a case-by-case basis (e.g., when calculating predictions in clinical practice). Multiple imputation methods that have widely been studied approximate the observed data using a well-known multivariate probability distribution (**section 4.1**) or approximate this distribution through a series of conditional (often regression-based) models (**section 4.2**). Although these methods can greatly differ in operationalization and underlying assumptions, simulation studies have demonstrated that they generally achieve similar performance (20,39,124). Overall, (semi-)parametric imputation methods can reliably be used for inference and prediction, and tend to perform well in datasets with a limited number of variables. Caution is, however, warranted when complex relations exist in the data (e.g., presence of treatment-covariate interactions), when observations are not independent (e.g., presence of repeated measurements) or when mechanisms of missingness are complex (e.g., presence of MNAR). In these situations, the required complexity of imputation methods drastically increases and manual configuration is often necessary to avoid bias (e.g., see **sections 6.1 and 6.2**) (37,41). In this regard, non-parametric methods offer several important advantages. First, there is no need to specify the functional form of the outcome relationship. Instead, non-linear effects and interactions are directly derived from the observed data. Second, there is no need to distinguish between different data types, as most machine learning methods can easily handle discrete, continuous and other data types. Third, because variable selection and dimensionality reduction are integrated into many machine learning procedures, they are well capable of dealing with high-dimensional datasets. Finally, because machine learning methods are extremely flexible, they are well suited to avoid incompatibilities between the imputation and substantive analysis model (41). This is an important issue when pursuing statistical inference and is often overlooked. Machine learning methods are therefore particularly appealing when there is limited understanding about likely sources of variability in the data, as data-driven procedures are used to determine how the imputations should be generated.

Results from the literature review are summarized in Figure 7. Each row in this figure represents one method of accommodating missing data, in order of appearance in this chapter. In particular, we highlight single imputation (single value imputation, expected value imputation), joint modelling imputation, conditional modelling imputation, non-parametric imputation, non-imputation methods, and methods dedicated for informative missingness and longitudinal data. Each cell displays the total number of simulation studies in which the method in the row outperforms the method in the column. Methods that work comparatively well have a higher percentage of papers in which they outperform other methods, signified by rows with many green cells. Most studies evaluated performance by quantifying the mean squared error of imputed values.

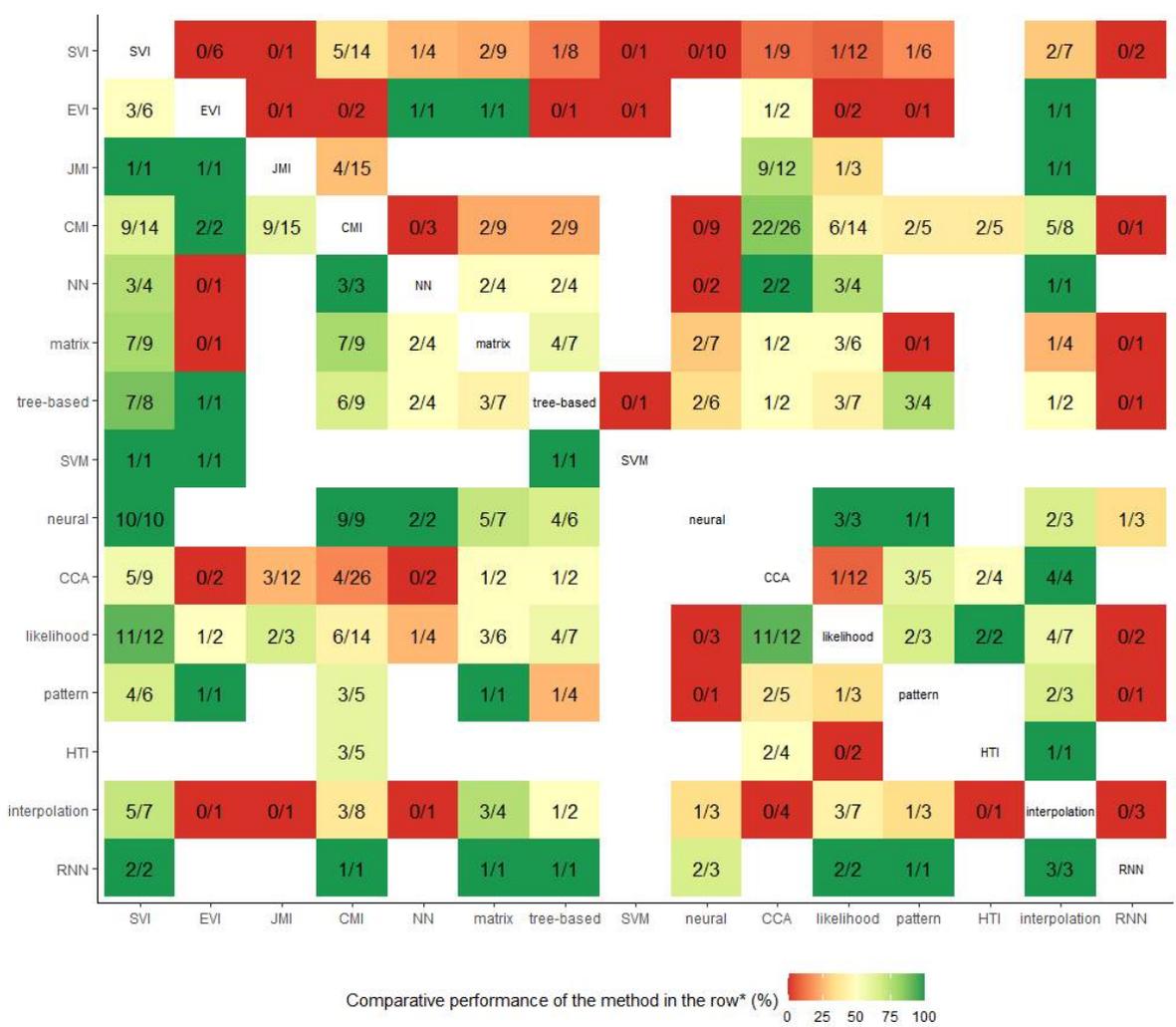

*Figure 9 Comparative performance of imputation methods as identified through a literature review. The color indicates the fraction of simulation studies in which the method in the row outperforms the method in the column.* **Single imputation methods:** *SVI = single value imputation; EVI = expected value imputation,* **Multiple imputation methods:** *JMI*

= joint modelling imputation, CMI = conditional modelling imputation, NN = nearest neighbor imputation, matrix = matrix factorization, tree-based = tree-based ensembles, SVM = support vector machine imputation, generative = neural network-based imputation, **Non-imputation methods:** CCA = complete case analysis, likelihood = likelihood-based approaches, pattern = missing data pattern methods, **Imputation of MNAR:** HTI = Heckman-type imputation, **Imputation of longitudinal data:** interpolation = interpolation methods (incl. last observation carried forward), RNN = recurrent neural networks.

Our literature review confirms that missing data is an important problem in RWD and requires dedicated methods. In particular, it is rarely justifiable to delete incomplete records, and to perform a so-called complete case analysis. Although missing values can accurately be recovered by adopting single imputation methods, simulation studies showed that their implementation usually leads to bias when estimating model parameters. For this reason, single imputation methods should be reserved for situations where imputations are needed on a case-by-case basis (e.g., when implementing a prediction model in clinical practice). Conversely, methods that perform consistently well are often based on multiple imputation using neural networks or other non-parametric approaches. As discussed, most of these methods can address mixed data types under various missingness mechanisms, and do not require user input to inform variable selection. Recurrent neural networks appear particularly useful because they can manage informative missingness and incomplete longitudinal data. However, when repeated measurements are relatively sparse, (semi-)parametric approaches that explicitly model their relatedness (e.g., through random effects) may be more suitable. Unfortunately, the implementation of multiple imputation methods can be very demanding w.r.t. available resources and may therefore not always be desirable. As discussed in **section 5**, it is possible to avoid the need for imputation in some circumstances. For instance, when adopting statistical models for prediction, the presence of missing data can simply be addressed by estimating pattern submodels (16). These models require fewer assumptions about the missing data mechanisms, and can perform well even when data are MNAR.

Finally, our review highlights several gaps in the published literature. Methods that appear promising but have not extensively been studied are based on SVM, or parametric models that estimate the joint distribution of the data and the missingness. Further, there is little consensus on appropriate strategies to evaluate missing data methods. For example, many simulation studies focus on situations where data are MCAR, or do not consider the validity of statistical inference that is based on imputed datasets. For this reason, it would be helpful to develop guidelines for the conduct and reporting of simulation studies focusing on missing data imputation, to facilitate fair comparisons between methods. For reasons of brevity, our review did not distinguish between different implementations of similar methods, such as the tree-based methods implemented within the chained equations framework. Uniting statistical and machine learning methods holds a promise to obtain imputations that are both accurate and confidence valid.

## Section 7. Summary

The analysis of RWD often requires extensive efforts to address data quality issues. In this chapter, we primarily focused on the presence of missing data and discussed several imputation methods. Although these methods are no panacea for poor quality RWD, their implementation may help address situations where RWD are incomplete or require recovery due to temporality or accuracy issues.

1. Assess whether missing data can be handled using non-imputation methods. For example, when the goal is to develop a prediction model, it is possible to avoid the need for imputation by adopting pattern submodels or built-in algorithms for dealing with missing values.
2. When pursuing imputation strategies, multiple imputed values should be generated to preserve uncertainty (and thus allow for inference)
3. Include the covariates and outcome from the substantive (analysis) model (78)
4. Include as many variables as possible, especially (auxiliary) variables that are related to the variables of interest or the presence of missingness (24,78).
5. Consider imputation methods that allow for informative missingness when missing data mechanisms cannot be ignored.
6. Especially in very large data sets with many cases and variables (RWD): use flexible imputation models (24). This can be achieved by adopting machine learning methods that have built-in procedures for variable selection and dimensionality reduction such as neural networks.
7. Evaluate the quality of imputed data by inspecting trace plots and distribution of imputed values (125)

## Acknowledgements

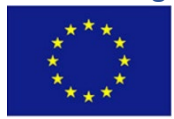 This project has received funding from the European Union's Horizon 2020 research and innovation programme under ReCoDID grant agreement No 825746.